\documentclass[aps,prl,twocolumn,superscriptaddress,showpacs]{revtex4}
\usepackage{graphicx}

\usepackage{dcolumn}

\usepackage{bm}
\usepackage{color}
\usepackage[colorlinks]{hyperref}
\input{epsf}
\begin{document}

% Use the \preprint command to place your local institutional report
% number in the upper righthand corner of the title page in preprint mode.
% Multiple \preprint commands are allowed.
% Use the 'preprintnumbers' class option to override journal defaults
% to display numbers if necessary
%\preprint{}

%Title of paper
\title{Radiation-induced oscillatory-magnetoresistance as a sensitive probe of the zero-field spin splitting in high mobility GaAs/AlGaAs devices}

\author{R. G. Mani}
\email{mani@deas.harvard.edu} \affiliation {Harvard University,
Gordon McKay Laboratory of Applied Science, 9 Oxford Street,
Cambridge, MA 02138, USA}

\author{J. H. Smet}
\author{K. von Klitzing}
\affiliation {Max-Planck-Institut f\"{u}r Festk\"{o}rperforschung,
Heisenbergstrasse 1, 70569 Stuttgart, Germany}

\author{V. Narayanamurti}
\affiliation {Harvard University, Gordon McKay Laboratory of
Applied Science, 9 Oxford Street, Cambridge, MA 02138, USA}
\affiliation {Harvard University, 217 Pierce Hall, 29 Oxford
Street, Cambridge, MA 02138, USA}

\author{W. B. Johnson}
\affiliation {Laboratory for Physical Sciences, University of
Maryland, College Park, MD 20740, USA}

\author{V. Umansky}
\affiliation{Braun Center for Submicron Research, Weizmann
Institute, Rehovot 76100, Israel}
%
% repeat the \author .. \affiliation  etc. as needed
% \email, \thanks, \homepage, \altaffiliation all apply to the current
% author. Explanatory text should go in the []'s, actual e-mail
% address or url should go in the {}'s for \email and \homepage.
% Please use the appropriate macro foreach each type of information
%
% \affiliation command applies to all authors since the last
% \affiliation command. The \affiliation command should follow the
% other information
% \affiliation can be followed by \email, \homepage, \thanks as well.
%\author{}
%\email[]{Your e-mail address}
%\homepage[]{Your web page}
%\thanks{}
%\altaffiliation{}
%\affiliation{}
%
%Collaboration name if desired (requires use of superscriptaddress
%option in \documentclass). \noaffiliation is required (may also be
%used with the \author command).
%\collaboration can be followed by \email, \homepage, \thanks as well.
%\collaboration{}
%\noaffiliation
%
\date{\today}
\begin{abstract}
We suggest an approach for characterizing the zero-field spin
splitting of high mobility 2D electron systems, when beats are not
readily observable in the Shubnikov-de Haas effect. The zero-field
spin splitting and the effective magnetic field seen in the
reference frame of the electron are evaluated from a quantitative
study of beats observed in radiation-induced magnetoresistance
oscillations
\end{abstract}
%
% insert suggested PACS numbers in braces on next line
\pacs{73.21.-b,73.40.-c,73.43.-f; Journal-Ref: Phys. Rev. B
\textbf{69}, 193304 (2004)}
% insert suggested keywords - APS authors don't need to do this
%\keywords{}
%
%\maketitle must follow title, authors, abstract, \pacs, and \keywords
\maketitle The recent discovery\cite{1} of novel non-equilibrium
zero-resistance states under photo-excitation might be counted as
the latest surprise in the physics of the 2-Dimensional Electron
System (2DES),\cite{1,2} which has provided topics of enduring
interest such as the integral and fractional quantum Hall effects
(QHE).\cite{3} Unlike quantum Hall systems, where zero-resistance
states coincide with a quantized Hall effect,\cite{3,4} these new
states are characterized by an ordinary Hall effect in the absence
of backscattering.\cite{1,2} The remarkable
phenomenology\cite{1,2} has generated much excitement due to the
possibility of identifying a new mechanism for artificially
inducing vanishing electrical resistance, while obtaining a better
understanding of QHE plateau formation and zero-resistance in the
2DES.\cite{5,6}

While the reasons mentioned above have motivated theoretical
interest,\cite{6} one might wonder whether the phenomena might
also serve as an improved probe in a 2D-problem with a functional
interest. Here, we suggest that radiation-frequency-independent
beats observed in the microwave-induced magnetoresistance might
serve as a sensitive indicator of the zero-field spin splitting
and the associated spin-orbit "Zeeman magnetic field" in the 2DES.
The resulting technique could help to advance research involving
the measurement, control, and utilization of the effective
magnetic field that appears in the reference frame of the
electron, in the high mobility GaAs/AlGaAs 2DES.\cite{7,8}

Mobile 2D electrons can experience, in their rest frame, an
effective magnetic field that develops from a normal electric
field at the semiconductor heterojunction interface due to the
so-called Bychkov-Rashba effect.\cite{9} As this magnetic field
can, in principle, be controlled by an electrical gate, it has
been utilized in the design of novel spin based devices.\cite{8}
The Bychkov-Rashba term, proportional to the electron wavevector
$k$, is believed to account for most of the observed Zero-field
Spin Splitting (ZFSS) in the narrow gap 2DES.\cite{8} In the 2DES
realized in wider gap GaAs/AlGaAs, a Bulk Inversion Asymmetry
(BIA) contribution to the ZFSS, provides an additional effective
magnetic field as B $\longrightarrow$ 0.\cite{10,11,12,13,14}
Although theory has suggested that the BIA term is stronger than
the Bychkov-Rashba term in the GaAs/AlGaAs heterostructure
2DES,\cite{14} ZFSS in n-type GaAs/AlGaAs is not easily
characterized because its signature is often difficult to detect
using available methods. This has hindered spintronics research in
the GaAs/AlGaAs system, which provides the highest mobility 2DES.

Typical investigations of ZFSS in the 2DES look for beats in the
Shubnikov-de Haas (SdH) oscillations that originate from
dissimilar Fermi surfaces for spin-split bands.\cite{15,16,17}
However, the applicability of this approach requires the
observability of SdH oscillations at very weak magnetic fields. In
GaAs/AlGaAs heterostructures with two dimensional electrons, this
is not easily realized, even in high quality material. In
addition, other mechanisms can, in principle, provide similar
experimental signatures.\cite{16,17,18} Thus, a need has existed
for simple new methods of investigating the spin-orbit interaction
in the 2DES characterized by a small zero-field spin splitting, to
supplement the SdH, Raman scattering, Electron Spin Resonance
(ESR) and weak localization based approaches.\cite{16,19,20,21} In
this light, the approach proposed here has the advantages of
simplicity and improved sensitivity because the ZFSS ($\approx$ 20
$\mu$eV) is determined through a comparison of the spin splitting
with an easily tunable, small energy scale set by the photon
energy ($\approx$ 200 $\mu$eV), unlike, for example, the SdH
approach which relates the ZFSS ($\approx$ 20 $\mu$eV) to
differences between two (spin split) Fermi surfaces with energy of
order $E_{F}$ $\approx$ 10 meV in GaAs/AlGaAs.

%%%%%%%%%%%%%%%%%%%%%%%%%%%%%%%%%%%%%%%%%%%%%%
\begin{figure}
%h=here, t=top, b=bottom, p=separate figure page
\begin{center}\leavevmode
\includegraphics[scale = 0.25,angle=0,keepaspectratio=true,width=3in]{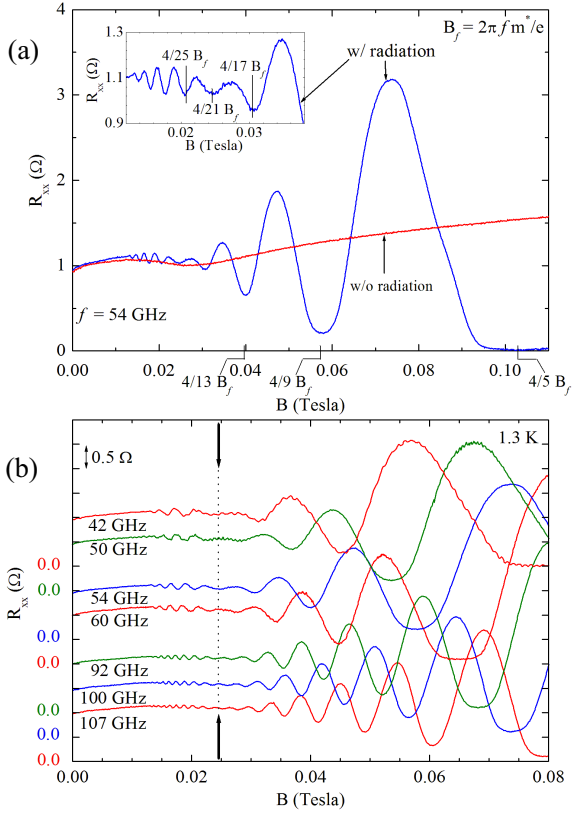}
\caption{(color online)(a): The magnetoresistance $R_{xx}$ in a
GaAs/AlGaAs heterostructure with (w/) and without (w/o) microwave
excitation. With radiation, $R_{xx}$ minima occur about $B$ =
$[4/(4j+1)] B_{f}$, and follow this empirical rule through the
beat, see inset. (b) $R_{xx}$ oscillations are shown for a set of
$f$. The beat remains fixed at B $\approx$ 0.024 Tesla as the
$R_{xx}$ oscillation frequency changes with $f$.}
\label{figurename1}\end{center}\end{figure}

%%%%%%%%%%%%%%%%%%%%%%%%%%%%%%%%%%%%%%%%%%%%%%%

Experiments were carried out on standard GaAs/AlGaAs
heterostructure devices. Upon slow cooling in the dark, the 2DES
exhibited a low electron density, $n$ (4.2 K) $\approx$
$1.5\times10^{11}$ $cm^{-2}$. Subsequent brief illumination by a
LED produced $n$(4.2 K) $\approx$ $3\times10^{11} cm^{-2}$ and a
mobility $\mu$(1.5 K) up to $1.5\times10^{7} cm^{2}/Vs$. Lock-in
based four-terminal measurements of the resistance and the Hall
effect were carried out with the sample mounted inside a
waveguide, as it was excited with microwaves over the frequency
range 27 $\leq$ $f$ $\leq$ 120 GHz. The microwave power, estimated
to be less than 1 mW, was set at the source and then reduced using
variable attenuators.

%%%%%%%%%%%%%%%%%%%%%%%%%%%%%%%%%%%%%%%%%%%%%%%
\begin{figure}
\begin{center}
\includegraphics*[scale = 0.25,angle=0,keepaspectratio=true,width=3in]{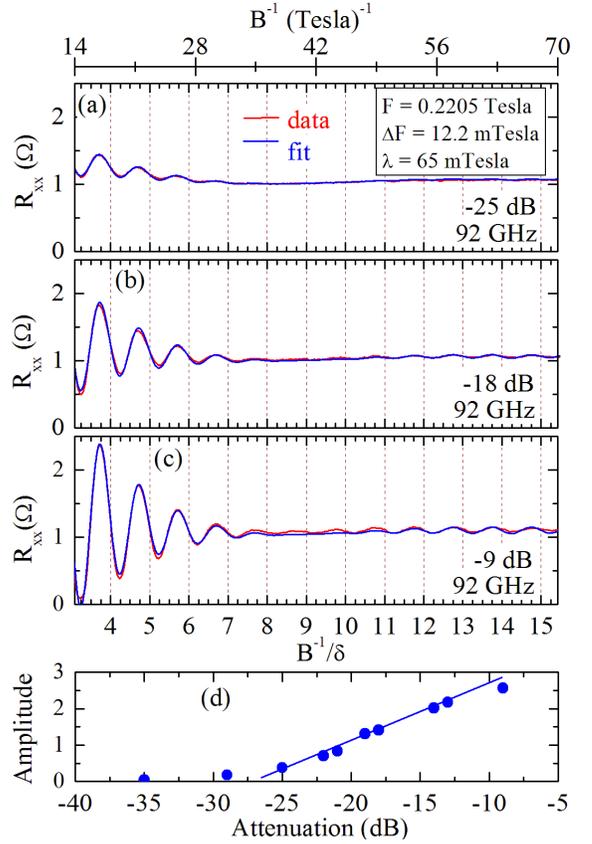}
\caption{(color online)(a)-(c) $R_{xx}$ oscillations are exhibited
vs. $B^{-1}$ (top axis) and $B^{-1}$/$\delta$ (bottom axis) with
the power attenuation factor as the parameter. Note that $\delta$
is the period in $B^{-1}$, i.e., $\delta$ = $F^{-1}$, where $F$ is
determined through a fit (see text). (d) The fit-amplitude, $A$,
of the $R_{xx}$ oscillations is plotted vs. the power attenuation
factor in decibels (dB). }\label{1}
\end{center}
\end{figure}
%%%%%%%%%%%%%%%%%%%%%%%%%%%%%%%%%%%%%%%%%%%%%%%%%%%%%%%%%%%%%%%%

Fig. 1(a) illustrates the magnetoresistance $R_{xx}$ observed with
(w/) and without (w/o) microwave excitation, in the high mobility
condition. Without radiation, weak magnetoresistance is
characterized by the absence of the SdH effect to B = 0.11 Tesla
at 1.3 K (see Fig. 1(a)). Excitation of the specimen with
electromagnetic waves induces oscillations in
$R_{xx}$,\cite{22,23} and a zero-resistance-state over a broad
B-interval in the vicinity of 0.1 Tesla.\cite{1,2} Fig. 1 (a)
shows that the three deepest resistance minima occur about (4/5)
$B_{f}$, (4/9) $B_{f}$, and (4/13) $B_{f}$, where $B_{f}$ = 2$\pi
f m^{*}/e$. In addition, a non-monotonic variation in the
amplitude of the resistance oscillations produces a beat at low
$B$ (see Fig. 1(a), inset).\cite{2} In Fig. 1(b), resistance data
are shown for a number of microwave frequencies. These data show,
for the first time, that the beat in the oscillatory resistance
remains at a fixed $B$ with a change in $f$.
%%%%%%%%%%%%%%%%%%%%%%%%
\begin{figure}
\begin{center}
\includegraphics*[scale = 0.25,angle=0,keepaspectratio=true,width=3.25in]{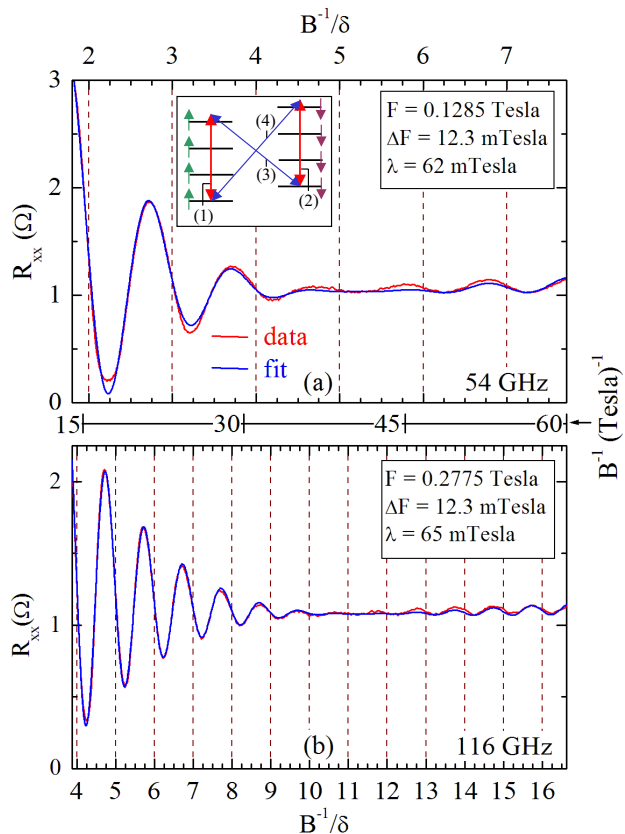}
\caption{(color online)$R_{xx}$ oscillations are shown at (a) 54
GHz and (b) 116 GHz, along with a fit (see text). The bottom- and
top- axis show $B^{-1}$/$\delta$, where $\delta$ = $F^{-1}$. Inset
to (a): The lineshape-fit includes the contributions cartooned
here. As $B$ $\longrightarrow$ 0, a zero-field spin splitting
produces an energy shift between the spin-up and spin-down levels.
}\label{1}
\end{center}
\end{figure}
%%%%%%%%%%%%%%%%%%%%%%%%%%%%%%%%%%%%%%%%%%%%%%%%%%%%%%%%%%%%%%%%

A lineshape study was carried out in order to characterize these
oscillations. Over a narrow $B$-window above the beat, the data
could be fit with an exponentially damped sinusoid: $R_{xx}^{osc}$
= $A'$ $exp(-\lambda/B)$$sin(2\pi F/B - \pi)$, where $A'$ is the
amplitude, $\lambda$ is the damping factor, and $F$ is the
resistance oscillation frequency.\cite{22} A lineshape that
included the superposition of two such waveforms: $R_{xx}^{osc}$ =
$A' exp(-\lambda/B)[sin(2\pi F_{1}/B - \pi) + sin(2\pi F_{2}/B -
\pi)]$ = $A exp(-\lambda/B) sin(2\pi F/B - \pi) cos(2\pi \Delta
F/B)$ where $F$ = $(F_{1}+F_{2})/2$ and $\Delta F$ =
$(F_{1}-F_{2})/2$, proved unsatisfactory in modelling the beat
data because this lineshape includes a phase reversal through the
beat, unlike experiment. Thus, we consider $R_{xx}^{osc}$ = $A
exp(-\lambda /B)$ $sin(2\pi F/B - \pi) [1 + cos(2\pi \Delta
F/B)]$, which can realize beats without phase reversal.

%%%%%%%%%%%%%%%%%%%%%%%%%%%%%%%%%%%%%%%%%%%%%%%
\begin{figure}
\begin{center}
\includegraphics*[scale = 0.25,angle=0,keepaspectratio=true,width=3.5in]{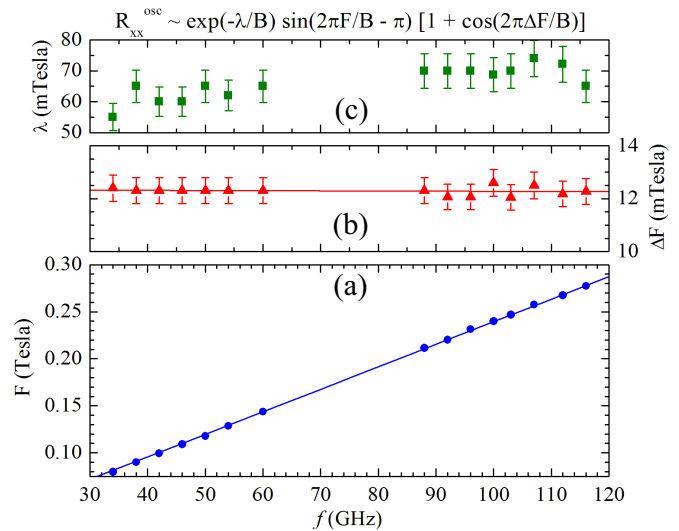}
\caption{(color online)Fit parameters obtained from this lineshape
study of the $R_{xx}$ oscillations. (a) The resistance oscillation
frequency, $F$, increases linearly with the radiation frequency.
(b) The beat frequency, $\Delta F$, is independent of $f$, and
$\Delta F$ $\approx$ 12.3 mTesla. (c) The damping parameter
$\lambda$ appears to be independent of $f$. }\label{1}
\end{center}
\end{figure}
%%%%%%%%%%%%%%%%%%%%%%%%%%%%%%%%%%%%%%%%%%%%%%%%%%%%%%%%%%%%%%%%

Fig. 2 (a) - (c) illustrate data obtained at a fixed $f$, as a
function of the power attenuation factor. Also shown, is a fit to
the data, using the lineshape $R_{xx}^{osc}$ = $A exp(-\lambda /B)
sin(2 \pi F/B - \pi) [1 + cos(2\pi \Delta F/B)]$. Inspection shows
good agreement between data and fit, and a comparison of Fig. 2
(a), (b), and (c) shows a monotonic increase in the amplitude of
the oscillations with increasing power level, that is reproduced
by the fit parameter, $A$, see Fig. 2(d). The oscillation period,
$\delta$, where $\delta$ = $1/F$, served to renormalize the
inverse field axis, as in the lower abscissa, see Fig. 2(c). The
data-plot vs. $B^{-1}$/$\delta$ shows that nodes occur in the
vicinity of $B^{-1}/\delta$ = $j$, and $B^{-1}/\delta$ =  $j +
1/2$, where $j$ =1,2,3,... while $R_{xx}$ minima transpire about
$B^{-1}/\delta$ = $[4/(4j+1)]^{-1}$.

Representative data and fit at a pair of $f$, see Fig 3, show that
this lineshape also describes the data obtained at widely spaced
$f$.  A summary of the fit parameters, $F$, $\Delta F$, and
$\lambda$, is presented in Fig. 4. The noteworthy features in Fig.
4 are: (i) $F$ increases linearly with $f$, (see Fig.
4(a)).\cite{1} (ii) The beat frequency, $\Delta F$ $\approx$ 12.3
mTesla, is independent of $f$ (see Fig. 4(b)), consistent with
Fig. 1(b). And, (iii) the damping parameter $\lambda$, $\lambda$
$\approx$ 65 mTesla, is also approximately independent of $f$ (see
Fig. 4(c)). Here, the exponential damping could be rewritten in a
Dingle form, $exp (-\lambda/B)$ = $exp(-\pi /\omega_{C}\tau_{f})$
$\approx$ $exp(-p T_{f} /B)$, where $T_{f}$ and $t_{f}$ represent
a finite frequency broadening-temperature and lifetime,
respectively, and $p$ is approximately 1. Note that $\lambda$ = 65
mTesla (Fig. 4(c)) corresponds to $T_{f}$  = 66 mK.\cite{22}

The lineshape, $R_{xx}^{osc}$ = $A exp(-\lambda /B) sin(2\pi F/B -
\pi) [1 + cos(2\pi \Delta F/B)]$, in our model can be understood
by invoking four distinct transitions between Landau subbands near
the Fermi level, as cartooned in the inset of Fig.3. Here, the
spin-orbit interaction helps to remove the spin degeneracy of
Landau levels as B $\longrightarrow$ 0. Thus, the terms (1) and
(2) represent spin preserving inter Landau level transitions, and
the terms (3) and (4) represent spin-flip transitions. If the
oscillations originating from these terms have equal amplitude and
share the same $\lambda$, then one expects a superposition of four
terms: $R_{xx}^{osc} = A' exp(-\lambda/B) [sin(2 \pi F/B - \pi) +
sin(2\pi F/B - \pi) + sin(2 \pi [F-\Delta F]/B - \pi) + sin(2 \pi
[F+\Delta F]/B - \pi)] = A exp(-\lambda/B) sin(2 \pi F/B - \pi) [1
+ cos(2\pi \Delta F/B)]$, which is the lineshape that has been
used here.

Here, the nearly equal amplitudes for the spin-flip and the
spin-preserving transitions are attributed to the occurrence of
mechanisms such as the Bychkov-Rashba effect and the
bulk-inversion symmetry,\cite{9,11} which produce spin precession
about an in-plane magnetic field.\cite{24} Under the influence of
such spin-orbit mechanisms, the effective magnetic field direction
experienced by electrons changes with scattering, and this
changing magnetic field environment helps to modify spin
orientation with respect to the applied magnetic field. A
component of the microwave magnetic field, which is oriented
perpendicular to the static magnetic field, can also serve to flip
spin.

Thus, beats observed in the radiation-induced resistance
oscillations appear as a consequence of a zero-field spin
splitting, due to the spin-orbit
interaction.\cite{8,9,10,11,12,13,14,15,16,17,19,20} One might
relate $\Delta F$ to the ZFSS by identifying the
radiation-frequency change, $\Delta f$, that will produce a
similar change in $F$, i.e., $\Delta f = \Delta F/[dF/df ]$, where
$\Delta F$ is the beat frequency, and $dF/df$  is the rate of
change of $F$ with the radiation frequency (see Fig. 4(a)). Then,
the ZFSS corresponds to $\Delta f$ = 5.15 GHz or $E_{S}(B=0)$ = $h
\Delta f$ = 21 $\mu$eV. From a study of the ESR at high magnetic
fields, Stein et al.\cite{20} reported a ZFSS of 7.8 GHz for $n$ =
$4.6 \times 10^{11}$cm$^{-2}$, which complements our result for
$n$ $\approx$ $3 \times 10^{11}$ cm$^{-2}$. Theory suggests that,
in the GaAs/AlGaAs heterostructure 2DES, both the BIA and the
Bychkov-Rashba terms have similar magnitudes, even as the BIA
makes the stronger contribution; the upper bound for the total
ZFSS is $\approx$ 80$\mu$eV.\cite{14}

Our ESR study near filling factor $\nu$ = 1 in the quantum Hall
regime showed that the electron spin resonance field, $B_{ESR}$,
varied as $dB_{ESR} /df$ = 0.184 Tesla/GHz, which implies an
effective Zeeman magnetic field $B_{Z}$ = [$dB_{ESR} /d f ] \Delta
f$ = 0.95 Tesla. As the effective magnetic field due to the
Bychkov-Rashba term is oriented perpendicular to both the
direction of electron motion and the normal of the 2DES, while the
BIA contribution lies along the direction of motion of
electrons,\cite{14} this evaluation of $B_{Z}$ is associated with
the magnitude of the vector obtained by adding these terms.
Although this estimate for the $B$ $\longrightarrow$ 0 Zeeman
magnetic field appears, at first sight, to be rather large by
laboratory standards, the small g-factor in GaAs/AlGaAs
heterostructures implies a small corresponding spin splitting,
$E_{S}$ = 21 $\mu$eV, in comparison to the Fermi energy ($\approx$
10 meV). Thus, SdH beats would be expected below the lowest B
($\approx$ 0.2 Tesla at 1.3 K) at which SdH oscillations were
observed in this material.

In summary, we have suggested that beats in the radiation-induced
magnetoresistance oscillations in GaAs/AlGaAs 2DES
heterostructures might serve as a new probe of the zero-field spin
splitting in the 2DES. This method could serve to simply track
changes in the spin-orbit interaction that result from the
controlled modification of the device structure, and this might
prove helpful for spintronics research based in the GaAs/AlGaAs
system.\cite{7,8}

 \vspace{0cm}


\begin{thebibliography}{24}
\bibitem{1} R. G. Mani, J. H. Smet, K. von Klitzing, V. Narayanamurti, W.
B. Johnson, and V. Umansky,  Nature (London) \textbf{420}, 646
(2002); Phys. Rev. Lett. \textbf{92}, 146801 (2004).

\bibitem{2} M. A. Zudov, R. R. Du, L. N. Pfeiffer and K. W. West, Phys. Rev. Lett.
\textbf{90}, 046807 (2003).

\bibitem{3} \textit{The Quantum Hall Effect}, 2nd Ed. edited by  R. E. Prange and S. M. Girvin,
(Springer-Verlag, New York, 1990).

\bibitem{4} D. C. Tsui, H. L. Stormer, and A. C. Gossard, Phys. Rev. B \textbf{25}, 1405 (1982).

\bibitem{5} R. Fitzgerald, Phys. Today \textbf{56} (4), 24 (2003)

\bibitem{6}  J. C. Phillips, Sol. St. Comm. \textbf{127}, 233
(2003); A. C. Durst, S. Sachdev, N. Read, and S. M. Girvin, Phys.
Rev. Lett. \textbf{91}, 086803 (2003); A. V. Andreev, I. L.
Aleiner, and A. J. Millis, Phys. Rev. Lett. \textbf{91}, 056803
(2003); P. W. Anderson and W. F. Brinkman, cond-mat/0302129; J.
Shi and X. C. Xie, Phys. Rev. Lett. \textbf{91}, 086801 (2003); A.
A. Koulakov and M. E. Raikh, Phys. Rev. B \textbf{68}, 115324
(2003); F. S. Bergeret, B. Huckestein, and A. F. Volkov, Phys.
Rev. B \textbf{67}, 241303 (2003); V. Ryzhii, Sov. Phys. - Sol.
St. \textbf{11}, 2078 (1970).

\bibitem{7} S. A. Wolf, D. Awschalom, R. Buhrman, J. Daughton,
S. von Molnar, M. Roukes, A. Chtchelkanova, and D. Treger, Science
\textbf{294}, 1488 (2001).

\bibitem{8} S. Datta and B. Das, Appl. Phys. Lett. \textbf{56}, 665
(1990).

\bibitem{9} Y. A. Bychkov and E. I Rashba, J Phys. C. \textbf{17}, 6039
(1984).

\bibitem{10} G. Dresselhaus, Phys. Rev. \textbf{100}, 580 (1955).

\bibitem{11} L. M. Roth, Phys. Rev. \textbf{173}, 755 (1968).

\bibitem{12} G. Lommer, F. Malcher, and U. Rossler, Phys. Rev. Lett. \textbf{60}, 728
(1988); P. Pfeffer and W. Zawadzki, Phys. Rev. B \textbf{59},
R5312 (1999); P. Pfeffer, \textit{ibid.} \textbf{59}, 15902
(1999).

\bibitem{13} R. Eppenga and M. F. H. Schuurmans, Phys. Rev. B. \textbf{37}, 10923 (1988).

\bibitem{14}E. A. De Andrada e Silva, G. C. La Rocca, and F. Bassani, Phys. Rev. B \textbf{50}, 8523 (1994).


\bibitem{15}B. Das, D. C. Miller, S. Datta, R. Reifenberger, W. P. Hong, P. K. Bhattacharya, J. Singh, and M. Jaffe,  Phys. Rev. B \textbf{39}, 1411 (1989).

\bibitem{16} P. Ramvall, B. Kowalski, and P. Omling, Phys. Rev. B \textbf{55}, 7160 (1997).

\bibitem{17} A. C. H. Rowe, J. Nehls, R. A. Stradling, and R. S. Ferguson, Phys. Rev. B \textbf{63}, 201307 (2001).

\bibitem{18} T. H. Sander, S. N. Holmes, J. J. Harris, D. K. Maude, and J. C. Portal, Phys. Rev. B 58, 13856 (1998).

\bibitem{19} B. Jusserand, D. Richards, G. Allan, C. Priester, and B. Etienne, Phys. Rev. B \textbf{51}, 4707 (1995).

\bibitem{20}D. Stein, K. von Klitzing, and G. Weimann, Phys. Rev. Lett. \textbf{51}, 130 (1983).

\bibitem{21} J. B. Miller, D. M. Zumbuhl, C. M. Marcus, Y. B. Lynda-Geller, D. Goldhaber-Gordon, K.
Campman, and A. C. Gossard, Phys. Rev. Lett. \textbf{90}, 076807
(2003).
\bibitem{22}R. G. Mani, J. H. Smet, K. von Klitzing, V. Narayanamurti, W. B. Johnson, and V. Umansky, Bull. Am. Phys. Soc. \textbf{46}, 972 (2001);
in \textit{Proceedings of the 26th International Conference on the
Physics of Semiconductors}, Edinburgh, 2002, edited by A. R. Long
and J. H. Davies, IOP Conf. Proc. No. \textbf{171},  (Institute of
Physics, Bristol, 2003) H112; cond-mat/0303034 (unpublished);
cond-mat/0305507 (unpublished).

\bibitem{23} M. A. Zudov, R. R. Du, R. R. Simmons, and J. L. Reno, Phys. Rev. B. \textbf{64}, 201311 (2001).

\bibitem{24}A. Khaetskii and Yu. Nazarov, Phys. Rev. B \textbf{61}, 012639 (2000)




\end{thebibliography}
\end{document}